\documentclass[sigconf]{acmart}
\renewcommand\footnotetextcopyrightpermission[1]{}
\def\BibTeX{{\rm B\kern-.05em{\sc i\kern-.025em b}\kern-.08emT\kern-.1667em\lower.7ex\hbox{E}\kern-.125emX}}

\setcopyright{none}
\makeatletter
\def\runningfoot{\def\@runningfoot{}}
\def\firstfoot{\def\@firstfoot{}}
\makeatother 

\begin{document}

\title{Emergent Leadership Detection Across Datasets}

\author{Philipp M\"uller}
\affiliation{%
\institution{Max Planck Institute for Informatics}
\institution{Saarland Informatics Campus}
}
\email{pmueller@mpi-inf.mpg.de}

\author{Andreas Bulling}
\affiliation{
  \institution{University of Stuttgart}
  \institution{Institute for Visualisation and Interactive Systems}
}
  \email{andreas.bulling@vis.uni-stuttgart.de}

\begin{abstract}

Automatic detection of emergent leaders in small groups from nonverbal behaviour is a growing research topic in social signal processing but existing methods were evaluated on single datasets -- an unrealistic assumption for real-world applications in which systems are required to also work in settings unseen at training time.
It therefore remains unclear whether current methods for emergent leadership detection generalise to similar but new settings and to which extent. 
To overcome this limitation, we are the first to study a cross-dataset evaluation setting for the emergent leadership detection task.
We provide evaluations for within- and cross-dataset prediction using two current datasets (PAVIS and MPIIGroupInteraction), as well as an investigation on the robustness of commonly used feature channels (visual focus of attention, body pose, facial action units, speaking activity) and online prediction in the cross-dataset setting.
Our evaluations show that using pose and eye contact based features, cross-dataset prediction is possible with an accuracy of 0.68, as such providing another important piece of the puzzle towards emergent leadership detection in the real world.

\end{abstract}

\maketitle

\section{Introduction}

\begin{figure}
\centering
  \includegraphics[width=1.0\columnwidth]{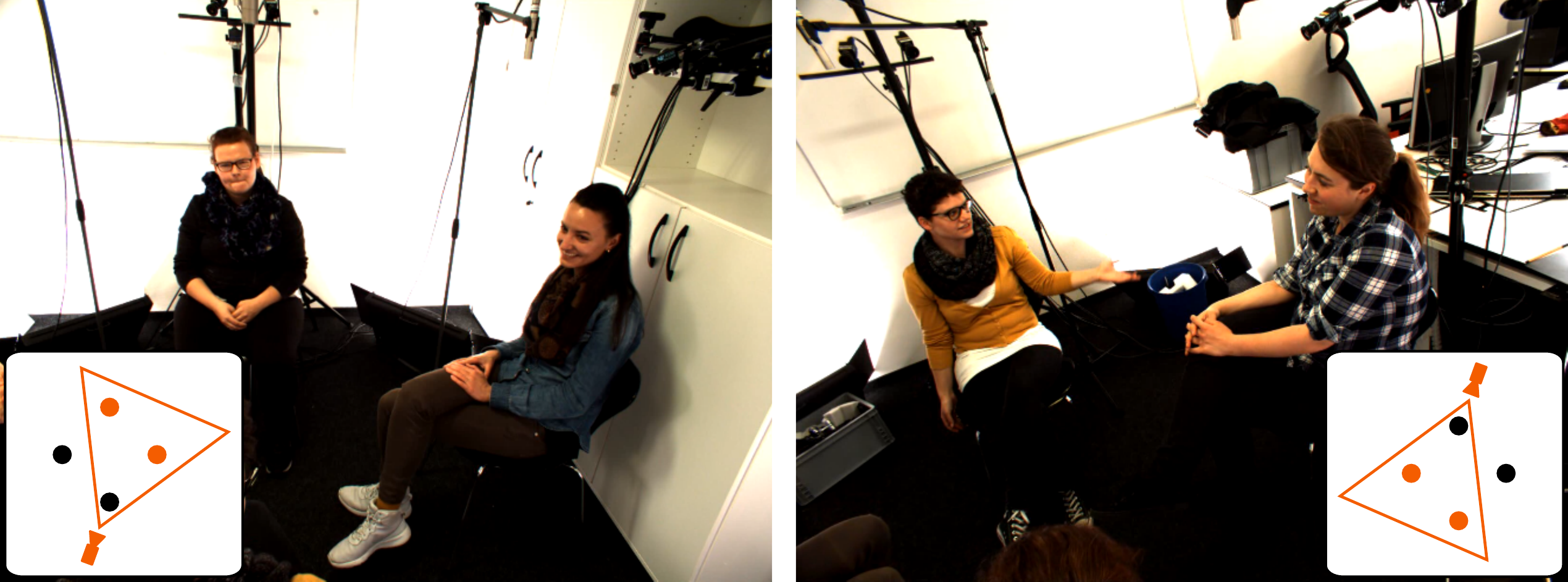}
  \caption{Illustration of the recording setup of the MPIIGroupInteraction dataset~\cite{mueller2018_iui}. The selected view and corresponding visible participants are shown in orange.\vspace{-2mm}}
  \label{fig:rec_setup}
\vspace{-2mm}
\end{figure}

Emergent leaders are group members who naturally obtain a leadership position through interaction with the group, and not via a higher authority~\cite{stein1979empirical}.
Even without formal authority, emergent leaders are important for group performance~\cite{druskat2006impact,kickul2000emergent}, and as a result
automatic identification of emergent leaders in group interactions is potentially beneficial in organisational research, for hiring decisions in the context of assessment centres~\cite{goodstein1999applications}, or for robots and intelligent agents that are supposed to interact with a group naturally.
Consequently, the detection of emergent leaders is a growing topic in social signal processing~\cite{feese2011discriminating,sanchez2012nonverbal,beyan2016detecting}.
These studies used nonverbal behaviour to detect emergent leaders in group interactions, which is supported by a large body of work on the connection between emergent leadership and nonverbal behaviour~\cite{baird1977some,gerpott2018eye,kalma1992gazing}.

While existent methods on emergent leadership detection in small groups showed reasonable performance, they all make the assumption that training and testing data come from the same distribution.
This assumption is unrealistic for application scenarios in which a system is required to detect emergent leaders in slightly different social situations for which no labelled data is available.
Until now, it remains unclear whether such cross-dataset leadership detection is possible with sufficient accuracy.

Specifically, emergent leadership detection in small groups of unaugmented people has only been investigated separately on two datasets employing very similar tasks, thereby ignoring the crucial cross-dataset setting.
The ELEA dataset ~\cite{sanchez2012nonverbal} consists of meetings of three or four people each, in which participants are given the winter survival task and instructed to come up with a joint solution.
Work on ELEA investigated emergent leadership detection from recordings of the meetings, by using audio- and visual or multi-modal features~\cite{sanchez2012nonverbal,sanchez2013emergent}, and more recently by using features obtained from a co-occurrence mining procedure~\cite{okada2019modeling}.
Kindiroglu et al. investigated domain adaptation and multi-task learning in order to predict leadership and extraversion on ELEA using video blogs annotated with personality impressions~\cite{kindiroglu2017multi}.
Their work is different to the cross-dataset setting described above, as they assumed access to emergent leadership ground truth on ELEA.

The PAVIS dataset~\cite{beyan2016detecting} consists of meetings of four people each either performing a ``winter survival task'' or a ``desert survival task''.
Research on the dataset focussed on detecting emergent leaders from nonverbal features only~\cite{beyan2016detecting}, using multiple kernel learning~\cite{beyan2016identification}, or using body pose based features~\cite{beyan2017moving}.
Further studies improved emergent leadership detection on the PAVIS dataset by using deep visual activity features~\cite{beyan2018investigation}, or by employing sequential analysis~\cite{beyan2019sequential}.
Apart from emergent leadership detection, the dataset has also been used to predict the leadership style of emergent leaders~\cite{beyan2018prediction,beyan2018investigation}.

Recently, M\"uller et al. recorded the MPIIGroupInteraction dataset of small group interactions to study low rapport detection~\cite{mueller2018_iui}.
Although emergent leadership ratings were recorded, no approach to leadership detection was proposed.
This dataset is particularly interesting for the emergent leadership detection task, as participants engaged in an open-ended discussion, which is in contrast to the rather constrained tasks that were performed on ELEA and PAVIS.

In this paper, we move one step closer to an emergent leadership detection system that can be applied in novel social situations without additional labelling effort.
We investigate emergent leadership detection across situations using two recent datasets~\cite{beyan2016detecting,mueller2018_iui} both featuring small group interactions but differing in participants' tasks, language, and nationality.
Our specific contributions are twofold:
We are the first to study emergent leadership detection in a cross-dataset setting, thereby achieving state-of-the-art results on MPIIGroupInteraction~\cite{mueller2018_iui}. 
Furthermore, we conduct extensive evaluations providing insights into the usefulness of different features and the feasibility of an online prediction system.

\section{Datasets}

To study cross-dataset emergent leadership detection, we utilise the PAVIS~\cite{beyan2016detecting} and the MPIIGroupInteraction~\cite{mueller2018_iui} datasets of small group interactions.

\subsection{PAVIS}

The PAVIS dataset~\cite{beyan2016detecting} consists of 16 interactions of four Italian speaking unacquainted participants of same gender each.
Each group performed either a ``winter-'' or a ``desert survival'' task, in which participants had to agree on a ranking of items they assume to be useful in a survival situation.
The interactions were recorded by four cameras, one facing each participant, 
as well as lapel microphones attached to each participant.
Interactions lasted from 12 to 30 minutes, resulting in a total corpus length of 393 minutes.
All recordings were divided into segments of four to six minutes and subsequently annotated for emergent leadership.
In line with previous work~\cite{beyan2018investigation}, we exclude four recordings due to audio problems, resulting in 12 meetings with a total of 48 participants.
We use PAVIS as a source dataset, as the segment-based annotation leads to more training data than is available on MPIIGroupInteraction~\cite{mueller2018_iui}.

\subsection{MPIIGroupInteraction}

MPIIGroupInteraction consists of 22 group interactions in German, each consisting of three- to four unacquainted participants.
In contrast to the rather constrained winter- or desert survival task on the PAVIS dataset~\cite{beyan2016detecting}, participants had an open-ended discussion.
The meetings were recorded by eight frame-synchronised cameras, two of them placed behind every participants in order to cover all other participants in their field of view (see Figure~\ref{fig:rec_setup}).
To record audio, one microphone was placed in front and slightly above participants' heads.
Each group was discussing for roughly 20 minutes, resulting in more than 440 minutes of audio-visual recordings in total.
After the interaction, each participant rated every other participant on a leadership scale (``PLead'' as in~\cite{sanchez2012nonverbal}).
We use the aggregate ratings for each participant to identify the ground truth emergent leader.

\section{Method}

After extracting nonverbal features from gaze, body pose, face and speaking activity we train Support Vector Machines to detect emergent leaders.

\subsection{Nonverbal Feature Extraction}

\begin{figure*}[t]
\centering
  \includegraphics[width=\textwidth]{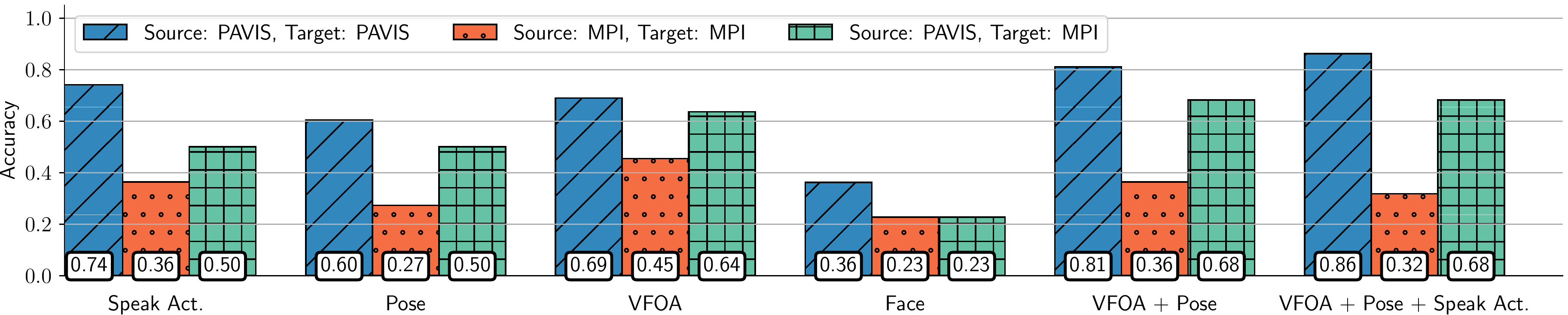}
  \caption{Performance of different featuresets when either training and testing on the same dataset, or training on PAVIS and testing on MPIIGroupInteraction. Random baseline for PAVIS as target is 0.25, for MPIIGroupInteraction as target 0.29.\vspace{-2mm}}
  \label{fig:results_offline}
\end{figure*}

\begin{figure}[t]
\centering
  \includegraphics[width=0.95\columnwidth]{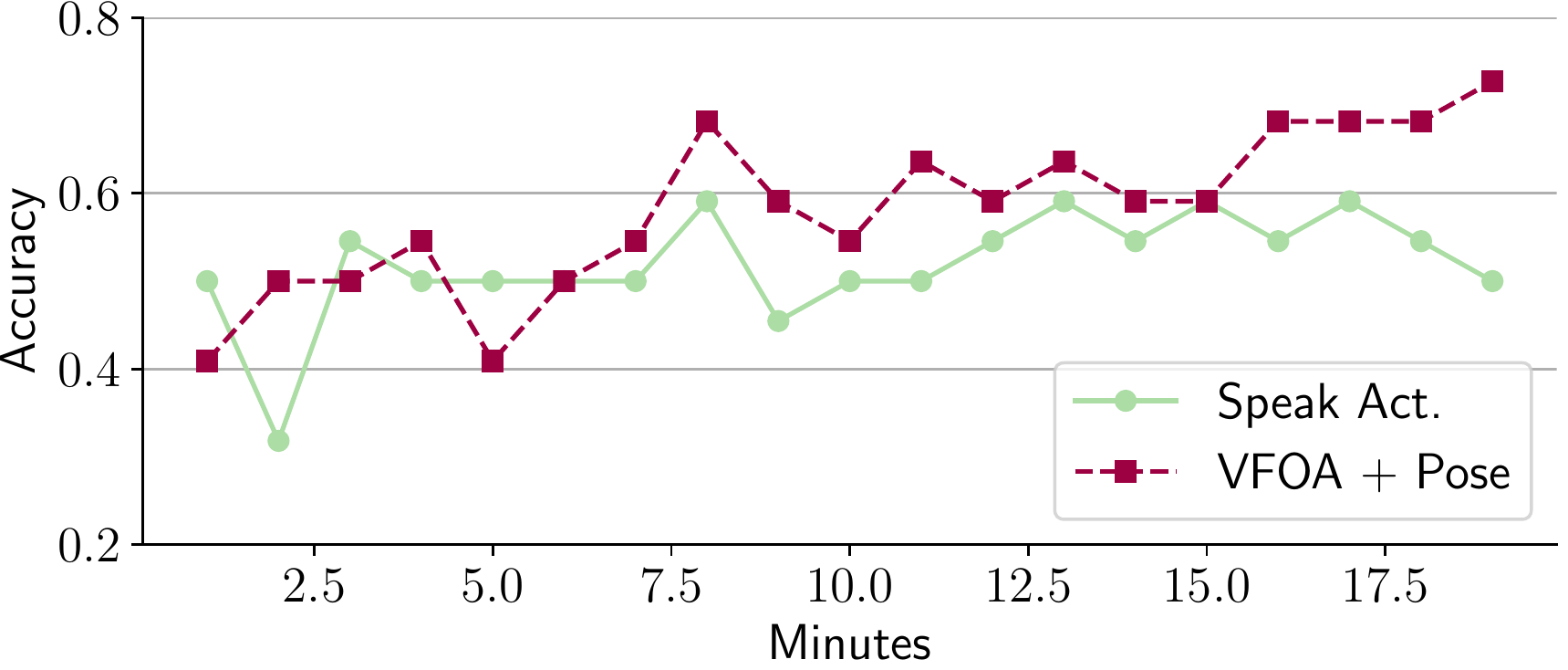}
  \caption{Performance of different featuresets when training on PAVIS and testing on MPIIGroupInteraction, depending on the size of the time window that is used for analysis (starting from the beginning). Random baseline is at 0.29.\vspace{-2mm}}
  \label{fig:results_extractionIntervalAnalysis}
\vspace{-2mm}
\end{figure}

\subsubsection{VFOA Features}

As the first step in computing features based on the visual focus of attention (VFOA), we perform eye contact detection, i.e. detecting at which other persons' face a target person is looking at a given moment in time.
To this end, we employ the recently introduced method for unsupervised eye contact detection in small group interactions by M\"uller et al.~\cite{mueller18_etra}.
To perform eye contact detection without the need of manual annotation, this method exploits the observation that people usually look at the person who is currently speaking in a weak labelling step.
The result are frame-wise predictions indicating with which other person the target person has eye contact, or whether the target person has no eye contact at all.
To arrive at optimal results, we use ground-truth annotated speaker segmentations as input to the method on MPIIGroupInteraction.
On PAVIS we resort to speaking activity detection via thesholding facial action units (cf.~\cite{mueller18_etra}), as we found the speaker segmentations provided with the dataset to not be perfectly synchronised with the video.
Using the eye contact annotations provided by the authors of~\cite{mueller18_etra} for evaluation, we obtain an accuracy of 0.7 on MPIIGroupInteraction.
To eliminate jitter, we apply a median filter of five frames to the eye contact predictions.

Based on these eye contact detections, we extract VFOA features as described in~\cite{beyan2016detecting}.
As the original implementation is not available from the authors, we implement the following features ourselves using the description in~\cite{beyan2016detecting}:
\textbf{totWatcher:} total time a person is watched by others,
\textbf{totME:} total time a person has mutual eye contact (MEC) with others,
\textbf{totWatcherNoME:} total time a person is being watched by others without having MEC,
\textbf{totNoLook:} total time a person is not looking at any other person,
\textbf{lookSomeOne:} total time a person looks at other people,
\textbf{totInitiatorME:} proportion of MECs of a person that are initiated by her,
\textbf{stdInitiatorME:} the standard deviation of lengths of MECs that are initiated by the person,
\textbf{totInterCurrME:} average time between intiation of a MEC and the start of the MEC,
\textbf{stdInterCurrME:} standard deviation of totInterCurrME,
\textbf{totWatchNoME:} total time a person is looking at others without MEC,
\textbf{maxTwoWatcherWME:} maximum time a person is looked at by two others,
\textbf{minTwoWatcherWME:} minimum time a person is looked at by two others,
\textbf{maxTwoWatcherNoME:} maximum time a person is looked at by two others without having MEC with them,
\textbf{minTwoWatcherNoME:} minimum time a person is looked at by two others without having MEC with them,
\textbf{ratioWatcherLookSOne:} ratio between totWatcher and lookSomeOne.
Note that while the features we compute on top on eye contact detections are the same as in~\cite{beyan2016detecting}, in the work by Beyan et al. they are based on VFOA detections using head pose.

\subsubsection{Body Pose Features}
We estimate body poses of participants using OpenPose~\cite{cao2018openpose} and follow the approach taken in~\cite{beyan2017moving} for pose feature computation.
In detail, we first detect frames of significant activity by a two-step thresholding approach on the difference images of subsequent greyscale frames:
In the first step a pixel is classified as moving if its value exceeds the threshold $T_1 = 30$ in the difference image.
The second step is classifying a frame as having significant activity if the number of moving pixels in it exceeds a threshold $T_2$.
We set $T_2$ such that we obtain the same proportion of frames with significant activity as described in ~\cite{beyan2017moving} (roughly 8.1\%).
For MPIIGroupInteraction we set $T_2$ for each interaction separately to not leak information between interactions at test time.

Subsequently, we compute the 80-dimensional featureset described in ~\cite{beyan2017moving} on frames with significant activity.
These features consist of statistical measures extracted from the angles between vectors that are defined by 2D joint positions. 
We use code provided to us by the authors of~\cite{beyan2017moving}.

\subsubsection{Facial Features}
We use OpenFace~\cite{baltrusaitis2018openface,baltruvsaitis2015cross} to extract presence and intensity of facial action units (AUs) following the approach described in~\cite{mueller2018_iui} for low rapport detection.
We specifically extract the means of AU activations and intensities and the mean and standard deviation of a ``facial positivity indicator''.

\subsubsection{Speaking Activity Features}
To evaluate the importance of speaking activity, we implement features used in previous work~\cite{sanchez2013emergent}, specifically the total speaking time of a participant (SPL), the number of speaking turns of a participant (SPT), the total number of times a participant interrupts other participants (SPI), and the average duration of a participants' speaking turns (ASP).
We normalise SPL, SPT and SPI with the length of the time interval from which we extract the feature.
On both datasets, we extract speaking activity features from ground truth speaker segmentations.

\subsection{Classification}

For classification, we use Support Vector Machines (SVMs) with radial basis function kernels. 
To obtain a single predicted leader for each interaction during test time, we obtain probability estimates using Platt scaling~\cite{platt1999probabilistic} and select the participant with the highest probability as the predicted emergent leader.
We choose the regularisation parameter $C$ of the SVM via cross-validation on the source dataset (PAVIS) and use the default value $1/n_{feats}$ for $\gamma$. 
Dedicated domain adaptation methods including Transfer Component Analysis~\cite{pan2011domain}, Correlation Alignment~\cite{sun2016return}, Random Walk Adaptation~\cite{van2017unsupervised} as well as transductive methods like label propagation~\cite{zhu2002learning} could not consistently improve over the plain SVM approach in our experiments.

The standard way to normalise both train and test data is via mean and standard deviation computed on the training data~\cite{friedman2001elements}.
This prevents information leakage from the test set at training time (e.g. when normalising train and test data jointly), and also leakage from ``future'' test samples at test time (when normalising the whole test set at once).
However, in our case training and testing data distributions differ and our data is structured by interactions made up of three to four individual participants.
As a consequence, while normalising the training data as usual, we normalise each test interaction separately (i.e. independently from the training data as well as other test interactions).
In this way, no information ``from the future'' is leaked while testing and we comply to the fact of different training and testing distributions.
In preliminary experiments, we found this way of normalising to be crucial.
The common way of normalising by applying mean and standard deviation of the training data on the test data resulted in much worse performance.

When employing several featuresets for classification, we always use late fusion, i.e. averaging scores of classifiers applied independently on the respective featuresets.
In preliminary experiments this produced more reliable results than early fusion.%

\section{Experimental Results}

All our evaluations are based on per-interaction accuracy of emergent leadership predictions as in~\cite{sanchez2012nonverbal,sanchez2013emergent}.
Specifically, an interaction is counted as correct, if and only if the predicted emergent leader is the same as the ground truth emergent leader.

\subsection{Offline Prediction}
To evaluate the extent to which classifiers trained on a source dataset are able to achieve high performance on a target dataset, we train on PAVIS and test on MPIIGroupInteraction.
At test time we assume to have access to a full test recording, i.e. we are predicting emergent leadership after an interaction took place (``offline'' setting).
In order to ensure using the same length for each of the approximately 20 minute long interactions on MPIIGroupInteraction we always use the first 19 minutes for feature extraction.

Figure~\ref{fig:results_offline} shows the obtained results for different feature sets and source- and target dataset combinations.
The highest performance in the cross-dataset setting (``Source: PAVIS, Target: MPI'') is achieved by a combination of VFOA and pose features with an accuracy of 0.68, slightly outperforming VFOA features only at 0.64 accuracy.
Combining other features with VFOA and pose did not improve results.
For applications where video recordings are not available or not desired, an accuracy of 0.5 can be achieved by relying on speaking activity features only.
Both results are clearly above the random baseline of 0.29, showing the feasibility of cross-dataset emergent leadership prediction.

Comparing cross-dataset to within-dataset results reveals that cross-dataset accuracies are consistently lower than within-dataset accuracies on PAVIS.
More surprisingly, by training on PAVIS, we achieve higher accuracies on MPIIGroupInteraction compared to training on MPIIGroupInteraction directly.
This is most likely an effect of the limited training data available on MPIIGroupInteraction.
In total there are only 78 samples (one per participant), whereas on PAVIS we have 232 samples due to the segment based annotations.

Concerning within-dataset results, we achieve the best accuracy for PAVIS with a combination of speaking activity, VFOA and pose features (0.86).
The best result for the emergent leadership detection task on PAVIS was published in~\cite{beyan2017moving}, achieving detection scores of 0.76 for the positive class and 0.93 for the negative class with a combination of pose and VFOA features.
Later work by the same authors adopted a different evaluation setting, and thus can not serve as a comparison~\cite{beyan2018investigation,beyan2019sequential}.
The detection scores for our predictions on PAVIS based on VFOA, pose and speaking activity features, reach 0.86 for the positive class and 0.95 for the negative class, exceeding the previously published results.
Within-dataset results on MPIIGroupInteraction are much lower, which is most probably due to the limited number of training examples.
Here, the best performance is reached by VFOA features at 0.45 accuracy.

\subsection{Online Prediction}

Some applications scenarios require information about emergent leaders already during the course of an interaction.
To evaluate performance in this setting, we restrict the time interval from which to extract features from the target interactions at test time.
Figure~\ref{fig:results_extractionIntervalAnalysis} shows accuracies for classifiers that only observe data from a limited number of minutes at the beginning of the interaction.
Both our best performing featureset (VFOA and pose) and speaking activity features tend to achieve higher accuracies after longer observation time.
This tendency is more pronounced for the VFOA and pose featureset, which stays between 0.4 and 0.6 accuracy during the first minutes of an interaction, and clearly above 0.6 accuracy after observing more than 15 minutes of the interaction.
Thus, while prediction above chance is possible early on, for optimal precision a significant portion of the interaction has to be observed.

\begin{table}[]
    \begin{tabular}{l | ll | ll}

        Feature                  & MPI& & PAVIS & \\ \hline
                                 &  Acc. & Ori. &  Acc. & Ori.\\ \hline
        totWatcherNoME           & 0.59    & $+$     & 0.66      & $+$      \\
        ratioWatcherLookSOne     & 0.59    & $+$     & 0.62      & $+$      \\
        totWatcher               & 0.55    & $+$     & 0.76      & $+$      \\
        totWatchNoME             & 0.55    & $-$     & 0.43      & $-$      \\
        totInitiatorME           & 0.45    & $-$     & 0.40      & $-$      \\
        lookSomeOne              & 0.45    & $-$     & 0.34      & $-$      \\
        stdInitiatorME           & 0.45    & $+$     & 0.34      & $+$      \\
        totNoLook                & 0.45    & $+$     & 0.34      & $+$      \\
        stdInterCurrME           & 0.45    & $-$     & 0.41      & $-$      \\
        maxTwoWatcherNoME        & 0.45    & $+$     & 0.21      & $+$      \\
        minTwoWatcherWME         & 0.45    & $-$     & 0.14      & $+$      \\
        maxTwoWatcherWME         & 0.41    & $+$     & 0.36      & $+$      \\
        minTwoWatcherNoME        & 0.41    & $-$     & 0.14      & $-$      \\
        totInterCurrME           & 0.41    & $-$     & 0.43      & $-$      \\
        totME                    & 0.36    & $+$     & 0.60      & $+$     

    \end{tabular}
    \caption{Accuracies for single feature based classification using VFOA features on PAVIS and MPIIGroupInteraction. ``Ori.'' indicates whether the maximum or the minimum of the feature was used for prediction.}
    \label{tab:single_feats}
\vspace{-2mm}
\end{table}

\subsection{Feature Analysis}

VFOA features were the best performing individual featureset in our evaluation.
To better understand which VFOA features generalise best across datasets, we investigate how well each individual feature discriminates the ground truth classes on MPIIGroupInteraction and PAVIS.
That is, for each interaction, we construct an unlearned classifier from a single feature by selecting the person with either the maximum or the minimum value on that feature as the emergent leader.
The choice of selection via maximum or minimum is based on achieved accuracy when comparing to ground truth.
We refer to features of which we take the maximum/minimum as having positive/negative orientation respectively.
It is important to note that this is not a valid classification approach, as we do not employ cross-validation. 
Instead, it is a post-hoc analysis on the connection between individual features and ground truth.
The results are summarised in Table~\ref{tab:single_feats}.
The usefulness of VFOA features for cross-dataset prediction is illustrated by the fact that all features except one share the same orientation.
The features with the highest accuracies on both datasets are \textit{totWatcherNoME}, \textit{ratioWatcherLookSOne} and \textit{totWatcher}.
This indicates that being looked at by others is a central property of leaders that is robust across datasets.
In contrast, the low performance of ~\textit{totME} on MPIIGroupInteraction in comparison to PAVIS indicates that mutual eye contact is less robustly associated with leadership across the two datasets.

\section{Conclusion}

In this paper, we are first to investigate a cross-dataset evaluation setting for the emergent leadership detection task.
We showed that it is possible to predict emergent leadership from nonverbal features on a new dataset that was not observed at test time.
We found that a combination of VFOA and pose features achieved best performance in the cross-dataset evaluation.
Furthermore, we analysed the feasibility of online prediction and the usefulness of single VFOA features.
All in all, our initial study on cross-dataset emergent leadership prediction opens the way to studying this important task in more realistic settings.

\begin{acks}
    This work was funded, in part, by the \grantsponsor{}{Cluster of Excellence on Multimodal Computing and Interaction (MMCI)}{} at Saarland University, Germany, as well as by a \grantsponsor{}{JST CREST}{} research grant under Grant No.:~\grantnum{1}{JPMJCR14E1}, Japan. We also thank Cigdem Beyan for sharing her code for pose feature computation with us.
\end{acks}

\bibliographystyle{ACM-Reference-Format}
\bibliography{references.bib}

\end{document}